\documentclass[proceedings]{JHEP3}
\PrHEP{PrHEP hep2001}			
\conference{International Europhysics Conference on HEP} 
\usepackage{epsfig}                   
\newcommand{\ben}{\begin{eqnarray}}   
\newcommand{\enn}{\end{eqnarray}}
\newcommand{\bea}{\begin{eqnarray}}   
\newcommand{\eea}{\end{eqnarray}}
\newcommand{\be}{\begin{equation}}
\newcommand{\ee}{\end{equation}}
\newcommand{\dst}{\displaystyle\phantom{\mid}}
\newcommand{\ov}{\over\displaystyle\strut}
\def\l({\left(}
\def\r){\right)}

\def\ave#1{\langle #1 \rangle}

\title{The reconstructed Big Bang from RHIC data
\footnote{Dedicated to J. Zim\'anyi on the occasion of his 70th birthday.}
}
\author{\speaker{Andr\'as Ster} \\		   
        MTA KFKI RMKI and MFA, H - 1525 Budapest XII,
        Konkoly-Thege 29-33, Hungary\\
        E-mail: \email{ster@mfa.kfki.hu}}

\author{Tam\'as Cs\"org\H{o}\\			   
        $^1$ Dept. Phys. Columbia University, 538 W 120th 
        St New York, NY -10027, USA\\
        $^2$ MTA KFKI RMKI, H - 1525 Budapest XII,
        Konkoly-Thege 29-33, Hungary\\	
	E-mail: \email{csorgo@sunserv.kfki.hu}}

\abstract{
        The final state of $Au + Au$ collisions 
	at $\sqrt{s}=130$ AGeV at RHIC has been 
        reconstructed within the framework of the Buda-Lund hydro model,
        by performing a simultaneous fit to preliminary PHENIX and STAR 
        data on two-particle Bose-Einstein correlations and identified
	single particle spectra, and the Hubble constant is determined to be
	$H = \langle u_t \rangle = 0.77 \pm 0.09$.
         }

\begin{document}

  \section{Introduction}

	The reconstruction of hadronic final state from the
	measured single particle spectra and two-particle correlation
	functions is of great current research interest in high energy
	heavy ion collisions, in order to identify 
	one or more new phases of hot and dense hadronic matter
	in the collisions of the biggest available	
	nuclei at the highest available bombarding energies. 
	It has been expected~\cite{bertsch,rischke-plot}, 	
	that for a Bjorken type of initial
	condition~\cite{bjorken} 
	and for a long-lived, soft, transient Quark	Gluon
	Plasma phase, pions are evaporated from a predominantly
	longitudinally expanding, transversely almost expansionless
	fireball during a long period of time and this 
	signature of the QGP phase can be observed experimentally
	from the analysis of two-pion Bose-Einstein correlation functions. 

	An alternative approach is to reconstruct the hadronic final
	state from the measureable single-particle spectra and
	two-particle correlation functions.  From this reconstructed final state
	and the knowledge of the equation of state of hot and dense hadronic 
	matter (e.g. from lattice QCD calculations) one can,
	 in principle, reconstruct the initial state 
	of the reaction by running the 
	(relativistic) hydrodynamical equations backwards in time, and
	determine if this initial state had been in the 
	QGP phase or not. Here we report on such 
	a reconstruction within the framework of the Buda-Lund hydro model.

\section{Buda-Lund hydrodynamic model}
	The Buda-Lund hydro parameterization (BL-H) has been developed
	in refs.~\cite{3d,3dqm} to describe particle correlations and
	spectra in central heavy ion collisions at CERN SPS.

	We assume that high energy heavy ion collisions can be 
	interpreted within the core-halo picture~\cite{chalo}
	where the center of the interactions is the core which is
	surrounded by the decay products of long lived resonances,
	corresponding to the halo.
        The Buda-Lund hydro parameterization describes the core
	of the particle emitting region as a cylindrically symmetric, finite
        hydrodynamically expanding system~\cite{3d,3dqm}, with emission 
        function
\be
        S_{c}(x,p) \, d^4 x   =  {\dst  g \ov (2 \pi)^3} \,
        \, { p^\mu d^4\Sigma_\mu(x) \ov
        \exp\l({\dst  u^{\mu}(x)p_{\mu} \ov  T(x)} -
        {\dst \mu(x) \ov  T(x)}\r) + s},
        \label{e:s}
\ee 
        where the subscript $_c$ refers to the core of collision
        (surrounded by a halo of long lived resonances).
	The degeneracy factor is denoted by $g$, the four-velocity field is
denoted by $u^{\mu}(x)$, the temperature field is denoted by $T(x)$,
the chemical potential distribution by $\mu(x)$ and $s = 0$, $-1$ or
$1$ for Boltzmann, Bose--Einstein or Fermi--Dirac statistics.  

The particle flux over the freeze-out layers is given by a generalized
Cooper--Frye factor, assuming that the freeze-out hypersurface depends
parametrically on the freeze-out time $\tau$ and that the probability
to freeze-out at a certain value is proportional to $H(\tau)$,
\bea
         p^\mu d^4\Sigma_\mu(x) & = &
         m_t \cosh[\eta - y]                                                   
        H(\tau) d\tau \, \tau_0 d\eta \, dr_x \, dr_y,\\
	H(\tau) & = & {(2 \pi \Delta\tau^2)^{-1/2}}
        \exp\left[-{(\tau - \tau_0)^2} / 
		({2  \Delta \tau^2}) \right],
\eea
	where we keep only the mean and the variance of the
	effective proper-time distribution $H(\tau)$.
        The transverse mass and coordinate are
	 $m_t = \sqrt{m^2 + p_x^2 + p_y^2}$ and
        $ r_t = \sqrt{r_x^2 + r_y^2}$,
        the rapidity $y$ and the space-time rapidity $\eta$ are
        defined as $y = 0.5 \log\left[(E+ p_z) / (E - p_z)\right]$ and
        $\eta = 0.5 \log\left[(t+ z) / (t - z)\right]$.
	
	The distributions of $1/T(x)$ and $\mu(x)/T(x)$ are 
	parameterized within BL-H  as
\ben
        {\dst \mu(x) \ov T(x) } & = & {\dst \mu_0 \ov T_0} -
        { \dst r_x^2 + r_y^2 \ov 2 R_G^2}
        -{ \dst (\eta - y_0)^2 \ov 2 \Delta \eta^2 }, \label{e:mu-r} \\
        {\dst 1 \ov T(x)} & =  &
        {\dst 1 \ov T_0 } \,\,
        \left( 1 + \Big\langle {\Delta T \over T}\Big\rangle_r\,
                {\dst  r_t^2 \ov 2 R_G^2} \right) \,
        \left( 1 + \Big\langle {\Delta T \over T}\Big\rangle_t
             \, {\dst (\tau - \tau_0)^2 \ov 2 \Delta\tau^2  } \right).
\enn   
        The central temperature and chemical potential 
	at the mean freeze-out time are denoted by
        $T_0 = T(r_x = r_y = 0; \tau = \tau_0)$ and
	$\mu_0 = \mu(r_x = r_y = 0; \tau = \tau_0)$.
        With the surface temperature $T_r = T (r_x = r_y = R_G,
        \tau = \tau_0)$ and the temperature after freeze-out,
        $T_t = T(r_x = r_y = 0;
        \tau = \tau_0 + \sqrt{2} \Delta\tau)$,
        the relative transverse and temporal temperature decrease
        are introduced, 
        see refs. ~\cite{ster-cf98,3dqm,3d,csrev} for further details.
	The variation of the chemical potential in coordinate
	space is related to the finiteness of the density profile
	in the core.
	
	The four-velocity $u^\mu(x)$ of the expanding matter is assumed to 
	have the form\cite{3d,uli_l,ster-beier}:
\bea
        u^{\mu}(x) & = & \l( \cosh[\eta] \cosh[\eta_t],
        \, \sinh[\eta_t]  \frac{r_x}{r_t},
        \, \sinh[\eta_t]  \frac{r_y}{r_t},
        \, \sinh[\eta] \cosh[\eta_t] \r), \nonumber \\
        \sinh[\eta_t]   & = & \ave{u_t} r_t / R_G,
\eea
        where $R_G$ stands for the transverse geometrical radius of the source.
	Such a flow profile, with a
	time-dependent radius parameter $R_G$, was recently shown to be an
	exact solution of relativistic hydrodynamics of a
	perfect fluid at a vanishing speed of sound~\cite{biro}.   
	It turned out~\cite{csell,csellgen}, 
	that the flow field is a generalized Hubble flow and
	the average transverse flow at the geometrical radius is 
	formally similar to Hubble's constant that characterizes the
	rate of expansion in our Universe,
	$\ave{u_t} = \dot R_G = H$. This emphasizes the 
	similarity between the Big Bang of our Universe  
	and the Little Bangs of heavy ion collisions.

\subsection{Single particle spectra and two particle correlations}
The invariant single particle spectrum is obtained~\cite{3d,csrev} as
\bea
        N_1({\bf k}) & = &
                {\dst d^2 n\ov  2 \pi m_t dm_t\, dy  } \,  = \, 
                {\dst g \ov (2 \pi)^3} \, \overline{E} \, \overline{V} \,
        \overline{C}
        \, {
        1 \ov
        \exp\l({\dst  u^{\mu}(\overline{x})k_{\mu}
        \ov  T(\overline{x})} -
        {\dst \mu(\overline{x}) \ov  T(\overline{x})}\r) + s}.
        \label{e:bl-n1-h}
\eea                                                                           
	The correlation function is found in the binary source formalism
	~\cite{3dcf98,csrev} as:
\ben
        C_2({\bf k}_1,{\bf k}_2)
        & = &
        1 +
         \lambda_* \, \Omega(Q_{\parallel}) \,\,\,
                \exp\left(- Q_{\parallel}^2 \overline{R}_{\parallel}^2
                 - Q_{=}^2 \overline{R}_{=}^2
                 - Q_{\perp}^2 \overline{R}_{\perp}^2 \right).
\eea
The pre-factor $\Omega(Q_\parallel)$ of the BECF induces oscillations within
the Gaussian envelope as a function of $Q_\parallel$. This oscillating
pre-factor satisfies $0 \le \Omega(Q_\parallel) \le 1$ and $\Omega(0)
= 1$. 
In practice, the period of oscillations is larger than the corresponding 
Gaussian radius, so the oscillations are difficult to resolve.
The above invariant BL-H form of the two-particle correlation function can be 
equivalently expressed in the frequently used but not invariant
Bertsch-Pratt (BP) form in the LCMS frame~\cite{lcms}, 
within the $\Omega=1$ approximation: 
\bea
	    C_2({\bf k}_1,{\bf k}_2)   & = &
        1 +  \lambda_* \exp\left[ - R_{s}^2 Q_{s}^2 - R_{o}^2 Q_{o}^2
        - R_l^2 Q_l^2 - 2 R^2_{ol} Q_o Q_{l} \right], 
\eea
	where the dependence of the fit parameters on the 
	value of the mean momentum of the pair is suppressed.  
        The above formulas for the BECF and IMD, as were used in the
        fits, have been introduced in
        refs.~\cite{3d,3dqm,ster-cf98,na22}, and summarized recently
	in ref.~\cite{csrev}. We recommend this latter review paper
	for the formulas that relate the BL-H model parameters
	to the above forms for the spectra and correlation functions,
	in particular, eqs.~(84-105), (115-118) and (129-140) 
	of ref.~\cite{csrev}.

\section{Fitting preliminary STAR and PHENIX data on Au + Au 	
	at RHIC}

        Here, we reconstruct the space-time picture
        of particle emission in Au + Au collisions at RHIC
	within the BL-H framework, by fitting simultaneously 
	the PHENIX and STAR preliminary data
        on two-particle correlations and single-particle spectra as
	presented at the Quark Matter 2001 conference,
refs.~\cite{phenix-qm01-imd,phenix-qm01-hbt,star-qm01-imd,star-qm01-hbt}.
	 For a proper core-halo correction
	$\propto$ ${1 / \sqrt{\lambda_*}}$ 
	the experimental values of the intercept parameter
	$\lambda_* (y,m_t)$ have to be taken from 
	the measurements.
	In the lack of these $\lambda_*(y,m_t)$
	values in ref.~\cite{phenix-qm01-hbt},
	we have utilized their average $\lambda_*$ for 
	a core-halo correction when fitting the PHENIX spectra. 
	In particular,
	the following average values were used  for the
	various particle types:
	$\overline \lambda_*(\pi) = 0.39 \pm 0.14$,
	$\overline \lambda_*(K) = 0.80 $ (estimated from the
	NA44 data on kaon-kaon correlations at CERN SPS~\cite{na44-kk-prl},
	$\overline \lambda_*(\overline{p}) = 1$ (the fraction
	of long lived resonances that decay to anti-protons is neglected).
	In case of the STAR data, we have utilized the same values for
	$\overline \lambda_*(K) $  and
	$\overline \lambda_*(\overline p)$
	however, for pions 
	we have utilized the $\lambda_*(m_t)$ values  of 
	ref.~\cite{star-qm01-hbt}.
	Note also that we have performed
	the data analysis within the $\Omega =1$  approximation.
	Here we improve on our
	earlier results~\cite{bl-datong} by taking into account an
	$m_t$ dependent core-halo correction for the STAR spectra
	and correlations, and by fitting the absolute normalization
	of the single particle spectra in both experiments, properly
	utilizing the  fugacity and quantum statistical factors.
	This allows us to extract the chemical potencial in the
	center of the fireball, in contrast to our 
	earlier fits~\cite{bl-datong}
	where the absolute normalization of the particle 
	spectra and the central value of the chemical potential
	distribution were not yet determined. 
	Unique minima are found and a good $\chi^2$/NDF 
	is obtained for both data sets.
	Within errors, all the fit parameters remained the same
	as in ref.~\cite{bl-datong}, but $\chi^2$/NDF decreased
	slightly.  See Figs. 1 and 2 for an illustation.
        The hypothesis that
        pions, kaons and protons are emitted from the same hydrodynamical
        source is in a good agreement with all the fitted data.

{\small
\TABULAR[tb]{|l|rl|rl||rl|rl|rl|}{
\hline
                 BL-Hydro
                 & \multicolumn{2}{c|}{STAR}
                 & \multicolumn{2}{c||}{PHENIX} 
                 & \multicolumn{2}{c|}{Au+Au}
                 & \multicolumn{2}{c|}{Pb+Pb}
                 & \multicolumn{2}{c|}{h+p}
                 \\
\cline{2-11}
                 parameters
                 & \multicolumn{2}{c|}{preliminary}
                 & \multicolumn{2}{c||}{preliminary} 
                 & \multicolumn{2}{c|}{$\langle RHIC \rangle$}
                 & \multicolumn{2}{c|}{$\langle SPS \rangle$}
                 & \multicolumn{2}{c|}{SPS}
                 \\
\hline
$T_0$ [MeV]      & 144  &$\pm$ 5    & 139  &$\pm$ 5     & 142  &$\pm$ 4    & 139
  &$\pm$ 6   & 140 &$\pm$ 3 \\
$\langle u_t \rangle$
                 & 0.86 &$\pm$ 0.10 & 0.68 &$\pm$ 0.3   & 0.77 &$\pm$ 0.09 & 0.55
 &$\pm$ 0.06 & 0.20  &$\pm$ 0.07\\
$R_G$ [fm]       & 8.0  &$\pm$ 0.5  & 6.6  &$\pm$ 0.3   & 7.3  &$\pm$ 0.7  & 7.1
  &$\pm$ 0.2 & 0.88 & $\pm$ 0.13 \\
$\tau_0$ [fm/c]  & 8.9  &$\pm$ 0.5  & 7.9  &$\pm$ 0.3   & 8.4  &$\pm$ 0.4  & 5.9
  &$\pm$ 0.6 & 1.4 & $\pm$ 0.1 \\
$\Delta\tau$ [fm/c]
                 & 0.5 &$\pm$ 1.0  & 0.6  &$\pm$ 1.2   & 0.5  &$\pm$ 1.1  & 1.6
  &$\pm$ 1.5 & $\ge$ 1.3  & $\pm$ 0.3 \\
$\Delta\eta$     & 1.0  &$\pm$ 0.1  & 1.5  &$\pm$ 0.1   & 1.2  &$\pm$ 0.3  & 2.1
  &$\pm$ 0.4  & 1.36 &$\pm$ 0.02\\
$\langle {\Delta T \over T}\rangle_r$
                 & 0.09 &$\pm$ 0.01 & 0.04 &$\pm$ 0.01  & 0.06 &$\pm$ 0.03 & 0.06
 &$\pm$ 0.05  & 0.71 &$\pm$ 0.14\\
$\langle {\Delta T \over T}\rangle_t$
                 & 1.6  &$\pm$ 0.4  & 0.86  &$\pm$ 0.09   & 1.2  &$\pm$ 0.4  & 0.59
 &$\pm$ 0.38 & - & \\
\hline
$\mu_0^{\pi^-}$ [MeV]  & 0 & (fixed) &  0 & (fixed)    & 0 & (fixed)    & - &    & - & \\ 
$\mu_0^{K^-}$ [MeV]  & 46 &$\pm$ 11   &  - &    & - &    & - &    & - & \\ 
$\mu_0^{\overline{p}}$ [MeV]  & 300 &$\pm$ 18   &  376 & $\pm$ 38   & 338 & $\pm$ 28   & - &   & - &\\  
\hline
$\chi^2/$NDF     & \multicolumn{2}{c|}{32/54 = 0.59}
                 & \multicolumn{2}{c||}{46/58 = 0.79}
                 & \multicolumn{1}{r}{0.69}
                 & \multicolumn{1}{l|}{ }
                 & \multicolumn{1}{r}{1.20}
                 & \multicolumn{1}{l|}{ }
                 & \multicolumn{1}{r}{0.94}
                 & \multicolumn{1}{l|}{ }
                 \\
\hline}
{\label{tab:results}
Preliminary source parameters from simultaneous fittings of
preliminary RHIC data of  PHENIX and STAR
on  particle spectra and HBT radius parameters with
the Buda-Lund hydrodynamical model. The third column
indicates their average. The (average)
fit parameters are shown for Pb+Pb collisions at CERN SPS
~\cite{qm99} and for h+p collisions at CERN SPS
~\cite{na22}. 
}
}

\DOUBLEFIGURE
{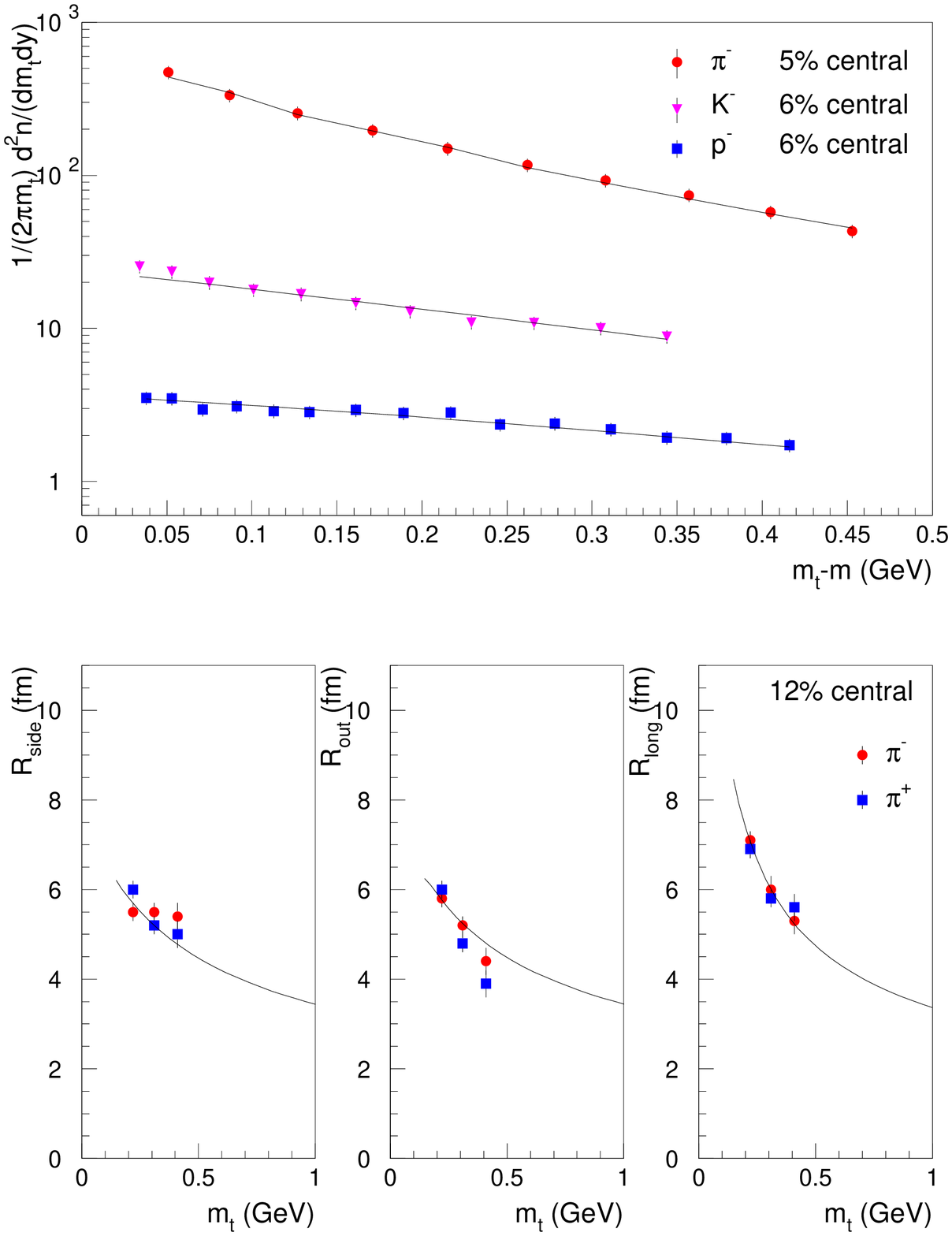,width=8.0cm}
{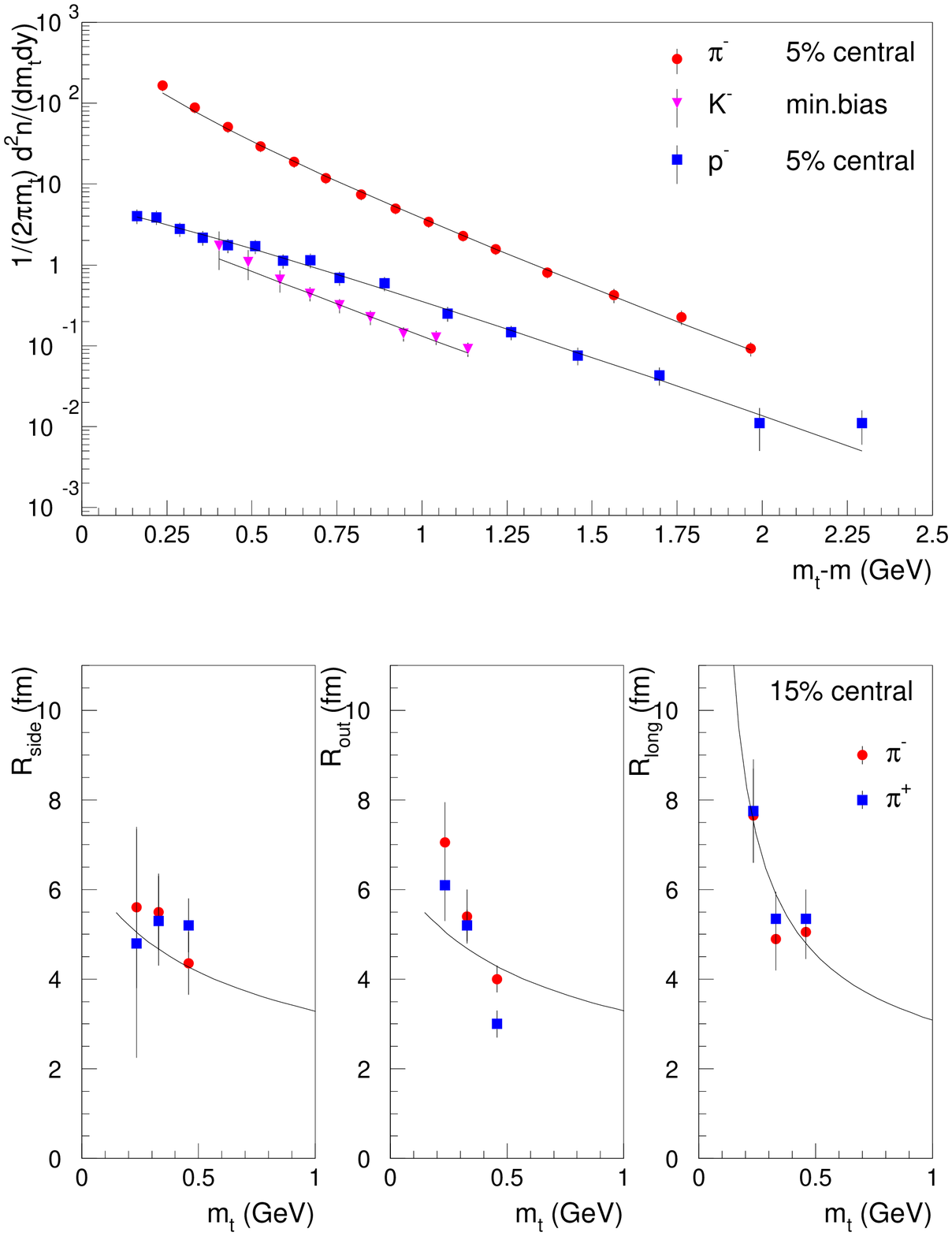,width=8.0cm}
{Simultaneous fits to STAR particle spectra and HBT radius parameters.
\label{figstar}}
{Simultaneous fits to PHENIX particle spectra and HBT radius parameters.
\label{figphenix}
\vspace{-0.5cm}}

\section{Conclusions}
        We find that the PHENIX and STAR data on single particle
        spectra of identified $\pi^-$, $K^-$ and $\overline{p}$ as well as 
        detailed $m_t$ dependent HBT radius parameters
        are consistent with the Buda-Lund hydro model as well as 
	with one another. The final state of central Au + Au 
        collisions at RHIC corresponds to a cylindrically symmetric, 
        large ($R_G = 7.3 \pm 0.7$ fm) and homogenous
        ($T_0 = 142 \pm 4 $ MeV) fireball. 
        A large mean freeze-out time, $\tau_0 = 8.4 \pm 0.5$ is found with 
        a short duration of particle emission and small inhomogeneities
	of the temperature profile during the particle emission process.
	At the reconstructed final state, the hadronic fireball
	expands  three-dimensionally with a strong transverse flow,
	characterized by the Hubble constant 
	$H = \langle u_t \rangle = 0.77 \pm 0.09$. 
	The major difference
	between the final state of heavy ion collisions
	at RHIC and at CERN SPS seems to be an increased freeze-out time
	and an increased transverse flow or Hubble constant at RHIC.
	Within the presently large errors we do not
	find a significant increase in the reconstructed
	geometrical source size when increasing the
	energy from SPS to RHIC.  This and other
	questions are being addressed by an attempt
	to include the pseudo-rapidity distribution at RHIC
	into the fitted data sets, and by repeating the analysis
	using final, published PHENIX and STAR data.

	We have found a non-vanishing chemical potential for 
	kaons and anti-protons
	in the center of the fireball from the absolutely normalized
	single-particle spectra, while the pion data were well described
	in all cases with a vanishing pion chemical potential 
	in the center of the fireball. These values together with
	the inhomogeneous chemical potential distribution
	of eq.~(\ref{e:mu-r}) indicate a clear deviation from
	chemical equilibrium in the reconstructed hadronic
	final state. 

	The similarities and the differences between an
	effective Quark Matter (QM) stage and a Quark Gluon Plasma (QGP)
	phase have been summarized recently in 
	ref.~\cite{csorgo-qm}. 
	The observed short duration of particle emission and the large
	transverse flow at RHIC contradicts to the picture of a 	
	soft, long-lived, evaporative Quark Gluon Plasma phase,
	that would consist of massless quarks and gluons.
	However, the final state does not exclude 
	a transient, explosive, suddenly
	hadronizing Quark Matter phase, that could be characterized by 
	massive valence quarks, the lack of gluons as effective 
	degrees of freedom, and a hard equation of state.

	This research has been supported by the Hungarian OTKA 
	T026435 and T034269, the NWO - OTKA grant N025186,
	by a NATO Science Fellowship, by an US NSF -
	Hungarian MTA-OTKA grant and by the DOE
	grant DE - FG02 - 93ER40764.

\end{document}